\documentclass[sigconf]{acmart}
\usepackage{array}
\usepackage{multirow}
\usepackage{graphicx}
\usepackage{makecell}
\usepackage{url}
\AtBeginDocument{%
  }

\copyrightyear{2024}
\acmYear{2024}
\setcopyright{rightsretained}
\acmConference[MM '24]{Proceedings of the 32nd ACM International Conference on Multimedia}{October 28-November 1, 2024}{Melbourne, VIC, Australia}
\acmBooktitle{Proceedings of the 32nd ACM International Conference on Multimedia (MM '24), October 28-November 1, 2024, Melbourne, VIC, Australia}
\acmDOI{10.1145/3664647.3681690}
\acmISBN{979-8-4007-0686-8/24/10}

\makeatletter
\gdef\@copyrightpermission{
  \begin{minipage}{0.3\columnwidth}
  \href{https://creativecommons.org/licenses/by-nc/4.0/}{\includegraphics[width=0.90\textwidth]{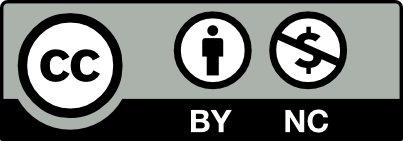}}
  \end{minipage}\hfill
  \begin{minipage}{0.7\columnwidth}
  \href{https://creativecommons.org/licenses/by-nc/4.0/}{This work is licensed under a Creative Commons Attribution-NonCommercial International 4.0 License.}
  \end{minipage}
  \vspace{5pt}
}
\makeatother

\begin{document}

\title{Dissecting Temporal Understanding in Text-to-Audio Retrieval}

\author{Andreea-Maria Oncescu}
\affiliation{%
  \institution{Visual Geometry Group,\\
University of Oxford}
\city{Oxford}
\country{United Kingdom}
 }
\email{andreea-maria.oncescu@sjc.ox.ac.uk}

\author{Jo{\~a}o F. Henriques}
\affiliation{%
  \institution{Visual Geometry Group,\\
University of Oxford}
\city{Oxford}
\country{United Kingdom}
 }
\email{joao@robots.ox.ac.uk}

\author{A. Sophia Koepke}
\affiliation{%
  \institution{T{\"u}bingen AI Center,\\ University of  T{\"u}bingen}
  \city{T{\"u}bingen}
 \country{Germany}
}
\email{a-sophia.koepke@uni-tuebingen.de}

\renewcommand{\shortauthors}{Oncescu, et al.}

\begin{abstract}
Recent advancements in machine learning have fueled research on multimodal tasks, such as for instance text-to-video and text-to-audio retrieval. 
These tasks require models to understand the semantic content of video and audio data, including objects, and characters. The models also need to learn spatial arrangements and temporal relationships. In this work, we analyse the temporal ordering of sounds, which is an understudied problem in the context of text-to-audio retrieval.
In particular, we dissect the temporal understanding capabilities of a state-of-the-art model for text-to-audio retrieval on the AudioCaps and Clotho datasets. Additionally, we introduce a synthetic text-audio dataset that provides a controlled setting for evaluating temporal capabilities of recent models. Lastly, we present a loss function that encourages text-audio models to focus on the temporal ordering of events. Code and data are available at \url{https://www.robots.ox.ac.uk/~vgg/research/audio-retrieval/dtu/}.
\end{abstract}




\maketitle

\section{Introduction}
The continued improvement of models and the increase in data available have led to impressive advances on various text-video tasks~\cite{Han_2024_CVPR,Kim_2024_CVPR,bain2022clip,Bain21,otani2020challengesmr,MDVC_Iashin_2020,Liu2019UseWY} and text-audio tasks~\cite{elizalde2023clap},
 including text-to-audio retrieval~\cite{Oncescu21a, koepke2022audio,flap2023,mei2023wavcaps,laionclap2023,wang2023one,chen23valor}, audio captioning~\cite{drossos2020clotho,Mei2021act,10448115} and recently, text-to-audio generation~\cite{kreuk2023audiogen,10112585,pmlr-v202-huang23i}. 
Understanding details, such as the temporal ordering of events, is important if we want our systems to give the best search results or generate reliable content for a text query. Recently, \cite{Wu2023AudioTextMD} showed that text-audio models do not use temporal cues available in text-audio datasets.\par

In this work, we build on \cite{Wu2023AudioTextMD} and examine the limitations of current state-of-the-art text-audio models, particularly in their use of temporal information. Different from \cite{Wu2023AudioTextMD} that considers a text-audio model containing a 
CNN-based audio encoder, our analysis uses the recent transformer-based audio encoder HTS-AT~\cite{htsat-ke2022} that serves as a component of state-of-the-art text-to-audio retrieval models~\cite{laionclap2023,mei2023wavcaps}.
We assess whether the model containing a transformer-based audio encoder results in better temporal understanding abilities than a CNN-based one. Additionally, we investigate the experimental designs and datasets used to determine if poor temporal understanding in current state-of-the-art models is caused by the training data or by the model architecture.\par

\begin{figure*}[t]
    \centering
    \includegraphics[width=\textwidth,height=0.25\textheight,keepaspectratio]{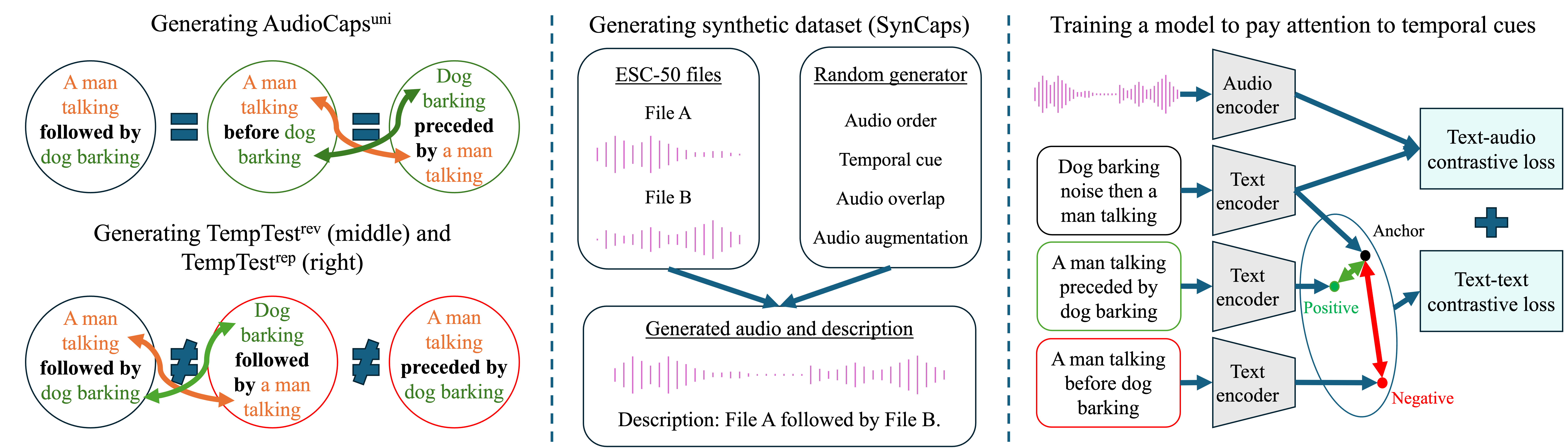}
    \caption{Analysing and improving the understanding of temporal cues in text-to-audio retrieval model. Top left:  AudioCaps$^{uni}$ is a variant of the AudioCaps dataset with modified descriptions that have a more uniform distribution of textual temporal cues. Additionally (bottom left), we generate test sets with  reversed temporal ordering or  replaced temporal conjunctions (TempTest$^{rev}$ and TempTest$^{rep}$). Middle: Generation of  audio-text pairs in the SynCaps dataset that provides a controlled evaluation setting.  Right: Our text-text contrastive loss improves temporal understanding using  positive descriptions (green) for the same sound ordering and negative examples (red) for the opposite temporal meaning.}
    \label{fig:summary_figure}
      \Description{Image split into three main pictures. One on the left shows circles containing similar or different text descriptions. The middle one contains 3 rectangles with text, showing how audio files and descriptions are being generated. The image on the right shows multiple shapes containing text and arrows.}
\end{figure*}

To determine whether commonly used text-audio datasets, such as AudioCaps~\cite{kim2019audiocaps} and Clotho~\cite{drossos2020clotho}, are suitable for training and evaluating the ability of current models to comprehend time, we examine the relative distribution of audio descriptions that contain temporal cues. In particular, we plot the frequency of specific temporal cues in relation to the total number of descriptions.
Our analysis shows that both the AudioCaps and Clotho datasets suffer from biases caused by the way humans describe events. That is, we tend to describe events in the order they appear. When first hearing the sounds of a dog barking and then the sound of a human speaking, we describe this as `A dog barking followed by a human speaking' rather than `A dog barking before a human speaks'. To try to address the lack of \textit{some} temporal examples, in \cite{Wu2023AudioTextMD}, the authors generate new text-audio pairs that enhance the existing text-audio data. They concatenate the audio files in a specific order and then generate a description that reflects that. For instance, if the generated sound is $Sound_1, Sound_2$, the description is `<Original description of $Sound_1$> before <Original description of $Sound_2$>'. This increases the size of the training data by 40\%. Different to \cite{Wu2023AudioTextMD}, we rephrase existing text descriptions to obtain a more uniform distribution of textual temporal cues whilst preserving the content (AudioCaps$^{uni}$). Furthermore, we investigate the impact of more uniform training data on the text-to-audio retrieval performance, reporting results on the original test data and on rephrased test data (TempTest$^{rev}$ and TempTest$^{rep}$). The rephrasing of text descriptions is illustrated in Fig.~\ref{fig:summary_figure}.\par

Furthermore, we present an empirical evaluation of the correctness and completeness of AudioCaps descriptions by leveraging a Large Language Model (LLM).
More specifically, we provide an LLM with (a subset of) AudioCaps descriptions and the start and end times of the sounds that make up the audio files (provided by~\cite{Xu2021TexttoAudioGB}). We ask the LLM to classify the sentences into correct -- if the description contains all the sounds and the correct ordering, incomplete -- if the description is missing sounds or is missing temporal context, and incorrect -- if the description contradicts the provided grounded sounds. We observe that about 23\% of the descriptions are incomplete or incorrect. This can contribute to models trained on AudioCaps not being able to understand temporal ordering.\par

To gain further insights into the temporal understanding capabilities, we propose a synthetic dataset that provides a controlled setting for analysing text-audio models. This dataset contains 10 second long audios to be consistent with the general setting that current models have been trained on.
We show that the considered model struggles to use temporal cues in the synthetic dataset, too, confirming the findings from \cite{Wu2023AudioTextMD} in a controlled setting. This allows us to decouple the bad temporal performance of the model from the data not being suited for the task.
Lastly, we propose a simple text-based contrastive loss function (see Fig.~\ref{fig:summary_figure}) and show that it results in the model paying more attention to the temporal ordering of events. This gives improvements in the overall retrieval results on the synthetic dataset.\par

In summary, we make the following contributions: (i) We show why existing text-to-audio retrieval datasets are not good indicators of a text-audio model's ability to understand temporal ordering, (ii) We propose a more uniform version of AudioCaps that is better suited for obtaining temporal understanding in models trained on this data. Additionally, its test subset allows for a better analysis of temporal understanding in existing models. This uniform version of AudioCaps keeps the audios intact and only requires changing the text descriptions. We provide benchmarks and an analysis of the behaviour of one current state-of-the art model on the original and more uniform versions of this dataset. (iii) We propose a synthetic dataset and use it to evaluate the model's understanding of time. (iv) We investigate an additional loss term to encourage the model to focus on text-based temporal cues.

\section{Related work}

\noindent\textbf{Text-to-audio retrieval.}
Text-to-audio retrieval involves matching a textual query with its most relevant audio file in a database of audio samples. The task of searching through audio databases can be approached in multiple ways. One simple approach is to match the text query with the title or the metadata of the audio file, provided it exists. However, for unlabelled databases, the aim is to find an audio file that has the content specified by the user through a text query. This is called semantic search. For many years, text-audio semantic retrieval has used audio class labels that consist of individual or few words as text queries~\cite{Wold96,slaney2002semantic, 10.1145/217279.215273, elizalde2019cross}. More recently, ~\cite{Oncescu21a,koepke2022audio} proposed new benchmarks where the text query is a free-form text description rather than a pre-defined class label, allowing for more control over the retrieved audio content. Collecting new text-audio pairs for training and using state-of-the-art transformer-based audio encoders has proven beneficial on the text-to-audio retrieval benchmarks~\cite{laionclap2023,mei2023wavcaps,flap2023}. As the annotation of audio files with descriptions is time consuming, some of the text-audio pairs collected by~\cite{laionclap2023} and~\cite{mei2023wavcaps} contain short audio labels instead of descriptions. To overcome this,~\cite{laionclap2023} employed the T5~\cite{T5} model to generate descriptions starting from audio labels, whilst~\cite{mei2023wavcaps} used ChatGPT~\cite{gptplayground}. \cite{mei2023wavcaps} also used ChatGPT to clean audio descriptions from datasets such as BBC Sound Effects\footnote{https://sound-effects.bbcrewind.co.uk/} by removing vision-based content. Another line of works considered metric learning objectives for text-to-audio retrieval~\cite{MeiXinhao2022OMLf, 10096972}. Other concurrent research pushed the text-to-audio retrieval results even further by training models with additional modalities, such as video and speech~\cite{chen2024vast,Wang2024InternVideo2SV}. Recently, \cite{oncescu2024sound} introduced new text-to-audio retrieval benchmarks on egocentric video data.

\noindent\textbf{Text-audio grounding.} \cite{Xu2021TexttoAudioGB} proposes a new set of data annotations for a subset of the AudioCaps dataset~\cite{kim2019audiocaps} with the aim of grounding each sound to a time interval. For this, annotators labelled the start and end times of all relevant sounds in each audio clip. \cite{Xu2024TowardsWS,10192960} investigated the task of weakly supervised text-to-audio grounding. The audio grounded dataset has also been used for learning to align sounds and text in an unsupervised manner~\cite{9747336}. \cite{audio_caption_metric} used the grounded sounds to introduce new metrics for audio captioning. In this work, we use this subset to provide an empirical evaluation of the quality of existing AudioCaps captions. More specifically, we give the grounded sounds and their corresponding AudioCaps descriptions to an LLM and ask it to evaluate if the descriptions are correct and complete.

\noindent\textbf{Temporal understanding in text-audio models.} \cite{Wu2023AudioTextMD} show that text-audio models do not pay attention to temporal cues in text queries, such as `followed by', or `after'. One example of an experiment in \cite{Wu2023AudioTextMD} is replacing temporal cues with words that represent a wrong ordering, e.g.\ replacing `then' with `as'. Then, the model's performance on the `wrong' descriptions is evaluated, revealing that this performance is similar to when the temporal ordering in the text queries is correct.
In their study,~\cite{Wu2023AudioTextMD} utilize CNNs for audio processing and identify a critical limitation of CNN-based models: applying temporal pooling across all embeddings can result in the loss of temporal information. To mitigate this issue, they augment the CNN architecture with several transformer layers to preserve temporal dynamics.
In contrast, contemporary models built on transformers inherently incorporate mechanisms to handle temporal data more effectively.
Different to \cite{Wu2023AudioTextMD}, we investigate the temporal understanding of a transformer-based state-of-the-art audio-text retrieval model. In particular, we analyse if a transformer-based model also ignores temporal cues. 
Additionally, the approach proposed by~\cite{Wu2023AudioTextMD} for helping models better understand time does not improve the overall performance on downstream retrieval benchmarks. In this work, we investigate an alternative for guiding the model to focus on temporal cues. Furthermore, we present a detailed analysis of descriptions in text-audio datasets in the context of temporal understanding. 
Concurrent work~\cite{Xu2024ADA} claims that commonly used text-audio datasets only contain simple audio descriptions that lack information about temporal cues, the number of times a sound can be heard, or details about sounds overlapping. To address this, \cite{Xu2024ADA} introduce a synthetic text-audio datasets by merging `atomic' sounds in a controlled way. 
Differently from~\cite{Xu2024ADA} that focuses on more varied audio details such as loudness, or number of times a sound can be heard, we  generate audio files and descriptions that showcase clear temporal relations between the composed sounds. In addition to that, \cite{Xu2024ADA} uses an LLM to generate descriptions which are varied but could include hallucinated content in contrast to the descriptions in our synthetic dataset which are rule-based.

\section{Analysis of temporal understanding in text-to-audio retrieval}\label{sec:audiocaps_analysis}
In this section, we take a close look at the AudioCaps and Clotho datasets in the context of understanding temporal information. We present a detailed evaluation of text-audio retrieval models on those datasets.

\subsection{AudioCaps dataset}\label{sec:audiocaps_dataset}
The AudioCaps~\cite{kim2019audiocaps} dataset contains paired audio clips and text descriptions. The training set consists of one text description for each audio file. The validation and test sets contain five descriptions for each audio file. 
In the AudioCaps evaluation setting, if at least one of the five text descriptions matches the audio clip, it counts as 100\% retrieval accuracy.

\begin{figure}
    \centering
    \includegraphics[width=0.9\columnwidth]{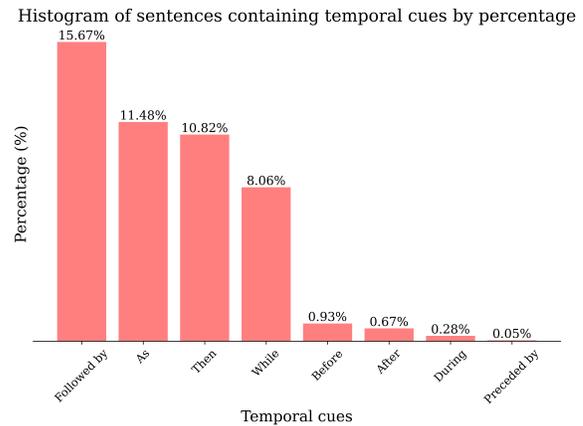}
    \caption{Distribution of temporal conjunctions and prepositions in the full AudioCaps~\cite{kim2019audiocaps} dataset. Most temporal sentences contain \textit{future}  temporal cues, such as `Followed by'. There is only a small proportion of \textit{past} cues, e.g.\ `Before'.}
      \Description{Pink histogram showing distribution of temporal cues as part of sentences in AudioCaps. Most sentences contain future and joint temporal cues.}
    \label{fig:all_splits_ac}
\end{figure}
\begin{figure}
    \centering
    \includegraphics[width=0.9\columnwidth]{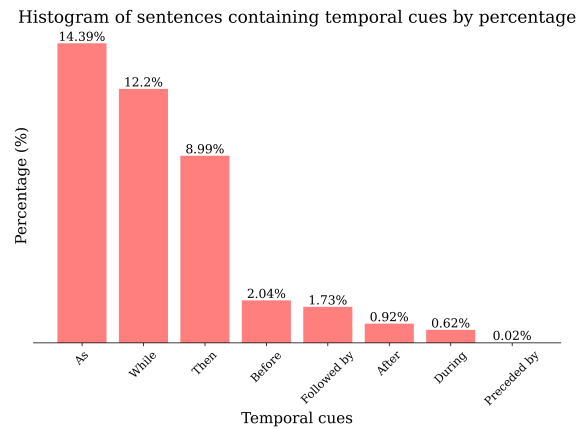}
    \caption{Distribution of temporal conjunctions and prepositions in the full Clotho~\cite{drossos2020clotho} dataset. Most temporal sentences contain \textit{joint} cues (e.g.\ `As', `While'), followed by \textit{future} ones (e.g.\ `Then'). Fewer sentences contain \textit{past} cues (e.g.\ `Before').}
      \Description{Pink histogram showing distribution of temporal cues as part of sentences in Clotho. Most sentences contain joint temporal cues followed by temporal cues that represent future events.}
    \label{fig:all_splits_clotho}
\end{figure}

The AudioCaps dataset is employed in all related text-to-audio retrieval works for training and evaluation. We want to understand the temporal characteristics of the AudioCaps dataset to gauge if the data available for training text-to-audio retrieval models is a part of the problem of models not understanding temporal cues~\cite{Wu2023AudioTextMD}.\par

First, we analyse the distribution of temporal conjunctions and prepositions in all the audio captions in the AudioCaps dataset in Fig.~\ref{fig:all_splits_ac}. 
We observe that most conjunctions and temporal prepositions suggest future events, i.e.\ `Followed by', `Then'. This is closely followed by the joint occurrence of audio events, i.e. `As' and `While'. However, almost no examples contain the temporal prepositions `Before' or `Preceded by' which is reasonable as humans would not naturally describe events in that order. A similar analysis is performed by~\cite{Wu2023AudioTextMD}, providing the distribution for `Followed by', `Then', `Before' and `After'. However, in~\cite{Wu2023AudioTextMD}, this distribution describes their training and test data which combines multiple datasets including AudioCaps~\cite{kim2019audiocaps} and Clotho~\cite{drossos2020clotho}. 
Here, we consider all words in the AudioCaps descriptions that represent temporal ordering. 
Given the distribution in Fig.~\ref{fig:all_splits_ac}, expecting a model trained on this data to understand the meaning of reverse temporal prepositions is unreasonable. At the same time, the test data also suffers from the same problem, therefore, using AudioCaps benchmarks for deciding if models understand temporal ordering is not optimal either. \par

Next, we \textit{empirically} evaluate the correctness and completeness of AudioCaps descriptions by using the grounded sound time intervals provided by~\cite{Xu2021TexttoAudioGB}. Through manual inspection, we notice that many AudioCaps descriptions are composed of multiple sounds. For instance, a 10 second audio file with a bird singing from second 0 to second 6 and a dog barking from second 4 to second 10 can be described as `Bird singing and/as dog barks'. Alternatively, this could be described as `Bird singing followed by dog barking'. Both descriptions are correct, however, a more complete version of these descriptions would be, for example, `Bird singing, soon joined by a dog barking. Their sounds overlap briefly before the bird stops, while the dog continues barking.'. If the description is not complete, however, how could a model learn the difference between `as' and `followed by' when they describe the same audio clip? \par

To \textit{empirically} evaluate the completeness and quality of the descriptions in AudioCaps with grounded sound sources, we use an LLM, specifically GPT-4~\cite{gptplayground}. We provide the LLM with the AudioCaps description, the grounded sources and their time intervals. We use one-shot prompting to give the model an example, such that it better understands the task. We then task the LLM to decide  if the AudioCaps description is  `correct', `incomplete', or `wrong' based on the sound source information.  Details for our prompt are shown in Tab.~\ref{tab:prompts_table}. To check how reliable the GPT-4 outputs are for this task, we manually checked 40 randomly selected descriptions, their grounded sounds and the GPT-4 evaluation. We found that the GPT-4 evaluation was correct 85\% of the time, with the majority of the mistakes being in favour of AudioCaps, i.e.\ evaluating a sentence as correct when in fact it was incomplete.
We show the resulting proportions of correct, incomplete, and wrong descriptions as identified by the LLM in the subset of AudioCaps in Tab.~\ref{tab:llm_eval}. On average, 23\% of the descriptions are incomplete or wrong. This percentage increases for descriptions containing \textit{future} and \textit{past} temporal cues.

The use of \textit{future} and \textit{past} refers to the fact that if `Sound 1' and `Sound 2' are connected by a \textit{future} temporal cue, then that means that `Sound 1' comes first and is followed by `Sound 2'. If a \textit{past} cue is used, then `Sound 1' comes after `Sound 2', e.g.\ `Bird sings after dog barks'. 
\textit{Future} cues include `Followed by', `Before' and `Then', e.g.\ `Bird sings before dog barks'. For \textit{past}, we consider `Preceded by' and `After'.
Based on the significant proportion of incomplete or wrong descriptions, and the distribution of temporal textual cues, we conclude that AudioCaps is not well-suited for analysing if text-audio models understand temporal ordering in audio clips and descriptions. \par
\begin{table*}
\caption{Methodology for evaluating the quality of the temporally grounded subset of AudioCaps using an LLM. Our input prompt includes setting the scene, one-shot prompting with an example for generating the evaluation of descriptions in the AudioCaps dataset, followed by the generation of new examples.}
\centering
\resizebox{0.96\textwidth}{!}{%
\begin{tabular}{p{6in}}
 \toprule
\multicolumn{1}{c}{Prompt} \\
\midrule
 Given descriptions of audio files and detailed temporal information about specific sounds within these files, where a sound may be present during multiple, distinct time intervals, your task is to evaluate the accuracy of each description with a primary focus on the timing and sequence of these sounds. Each audio file is 10 seconds long. For every description, assess its accuracy specifically in terms of how well it captures the chronological order and exact timing of sounds. Classify your evaluation into one of three categories: `Correct', `Incomplete', or `Wrong'. If necessary, provide a corrected description that not only fixes inaccuracies related to timing but also maintains the original writing style of the description. Your analysis should critically examine the temporal details provided, ensuring your assessment is primarily guided by the accuracy of these temporal sequences.\newline \newline Keep in mind the following:\newline Pay attention to whether the description matches the start and end times of sounds accurately.\newline Consider if the sequence of described sounds follows the actual sequence in the audio file. Evaluate if the description misses any sounds within the specified time frames or includes sounds that do not occur within these times.\newline Use similar vocabulary as the original audio description.\newline \newline Example:\newline Input:\newline Original audio description: A power tool motor running then revving\newline Localized components and their start and end times:\newline revving: 2.154, 10.02;\newline a power tool motor running: 0.0, 10.02;\newline Output:\newline Evaluation: Incomplete\newline Corrected description: A power tool motor running throughout, with revving starting early on and continuing alongside the motor's running sound until the end.\\
  \addlinespace

\bottomrule

\end{tabular}%
}
\label{tab:prompts_table}
\end{table*}
\begin{table}
\caption{Proportion of correct, incomplete and wrongly captioned AudioCaps data as determined by an LLM. First row contains the total numbers of temporally grounded descriptions. The other rows show proportions for specific temporal cues. Almost a quarter of temporal sentences are incomplete or wrong.}
\centering
\begin{tabular}{cccc}
 \toprule
  Preposition & Correct & Incomplete & Wrong \\
 \midrule
  Total (\#) & 3835 & 636 & 503\\
 \midrule
 As (\%) & 75.3 & 13.9 & 10.8 \\
 Followed by (\%) & 60.6 & 15.3 & 24.1 \\
 Then (\%) & 62.1 & 15.9 & 22.0 \\
 While (\%) & 72.0 & 15.4 & 12.6 \\
 Before (\%) & 58.8 & 9.8 & 31.4 \\
 After (\%) & 54.5 & 6.1 & 39.4 \\
 Proceeded by (\%) & 50.0 & 0.0 & 50.0 \\
 During (\%) & 53.3 & 13.3 & 33.3 \\
 And (\%) & 75.6 & 13.5 & 10.9\\
\bottomrule
\end{tabular}
\label{tab:llm_eval}
\end{table}

\subsection{Clotho dataset}
We examine whether the Clotho~\cite{drossos2020clotho}  dataset, another popular text-audio retrieval dataset, faces similar temporal limitations as AudioCaps. Clotho contains pairs of audio clips and text descriptions. Unlike AudioCaps, the train, validation, and test subsets contain five descriptions for each audio file. For a more uniform experimental setting, in the next sections, instead of pairs of one audio and 5 corresponding text descriptions, we split the Clotho training pairs into 5 pairs of one audio (repeated) and one corresponding text description. \par
We investigate the distribution of sentences containing temporal cues in Clotho. We notice in Fig.\ref{fig:all_splits_clotho} that, similarly to AudioCaps, Clotho contains a non-uniform distribution of temporal cues. However, for Clotho, descriptions mainly feature simultaneous actions, with the words `As' and `While' being used most often. This is followed by  \textit{future} temporal cues such as `Then' or `Followed by'. There are three times fewer sentences containing \textit{past} temporal cues. Therefore, we expect that models trained on Clotho are biased towards simultaneous or in-order (\textit{future}) events.

\subsection{Model performance on AudioCaps and Clotho}\label{sec:audiocaps_experiments}

In this section, we investigate the performance of a state-of-the-art model for text-to-audio retrieval on AudioCaps and Clotho in detail.

\subsubsection{Evaluation metrics.} Throughout all experiments, we use the standard evaluation metrics for retrieval:
recall at rank $k$ (R@$k$). This measures the percentage of targets
retrieved within the top $k$ ranked results. Higher numbers are better. We report results for text-to-audio (T $\rightarrow$ A) and audio-to-text retrieval (A $\rightarrow$ T).
We report the mean of three runs that use different random seeds.\par

\subsubsection{Model}
We employ the state-of-the-art text-audio model by~\cite{mei2023wavcaps}, utilising an HTS-AT audio encoder~\cite{htsat-ke2022}, and a pre-trained BERT encoder for text. After encoding audio and text inputs, an MLP projects the embeddings into the same space. We use the model variant pre-trained on the joint dataset of WavCaps~\cite{mei2023wavcaps} $+$ AudioCaps~\cite{kim2019audiocaps} $+$ Clotho~\cite{drossos2020clotho}. In our experiments, we finetune the model for 40 epochs on the dataset of interest (e.g. AudioCaps, Clotho) and use the same setup as \cite{mei2023wavcaps}. 
The best model is selected using the highest average validation retrieval accuracy R@1 on the dataset used for experiments. We run these experiments on A6000s.

\subsubsection{Loss function}
We use the same loss as \cite{mei2023wavcaps} - a normalised temperature scaled bidirectional cross-entropy loss (NT-Xent)~\cite{10.5555/3524938.3525087}. We refer to this as $\mathcal{L}_{at}$ with
\begin{equation}\label{eqn:sim}
s_{a_{i}t_{j}} = \frac{f(a_{i}) \cdot g(t_{j})}{\lVert f(a_{i}) \rVert_2 \lVert g(t_{j}) \rVert_2},
\end{equation}
\begin{equation}
\resizebox{.9\columnwidth}{!}{$
   \mathcal{L}_{ta} = -\frac{1}{2B} \sum_{i=1}^{B} \left[ \log \left( \frac{\exp(s_{a_{i}t_{i}}/\tau)}{\sum_{j=1}^{B} \exp(s_{a_{i}t_{j}}/\tau)} \right) + \log \left( \frac{\exp(s_{a_{i}t_{i}}/\tau)}{\sum_{j=1}^{B} \exp(s_{a_{j}t_{i}}/\tau)} \right) \right]
$}.
\end{equation}
Here $f(\cdot)$ is the audio encoder and $g(\cdot)$ the text encoder. $s_{a_it_j}$ is the cosine similarity, $a_i$ the audio input, $t_j$ the text input, $B$ the batch size, and $\tau$ is a temperature parameter. More details can be found in \cite{mei2023wavcaps}.

\subsubsection{Data used}
\begin{figure}
    \centering

    \includegraphics[width=0.98\columnwidth]{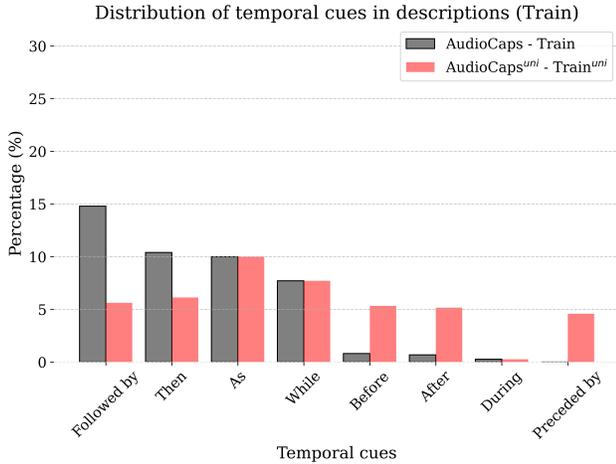}
    \caption{Distribution of temporal conjunctions and prepositions in the AudioCaps training data. We compare the proportion of temporal textual cues in the original training dataset (Train) and our proposed variant with a more uniform distribution of temporal textual cues ($Train^{uni}$).
    }
  \Description{Grey and red histogram with percentages shown on the vertical axis and words being shown on the horizontal axis.}
    \label{fig:train_temp_per}
\end{figure}

\begin{figure}
    \centering

    \includegraphics[width=0.98\columnwidth]{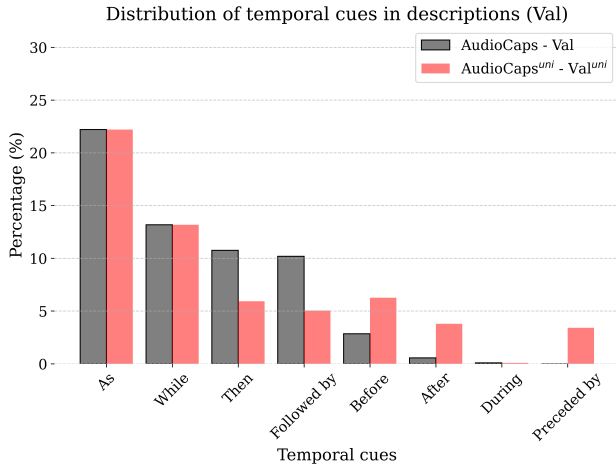}
    \caption{Distribution of temporal conjunctions and prepositions in the AudioCaps~\cite{kim2019audiocaps} validation dataset (Val) compared to our proposed variant with a more uniform distribution of temporal textual cues ($Val^{uni}$).
    }
  \Description{Grey and red histogram for the AudioCaps validation set with percentages shown on the vertical axis and words being shown on the horizontal axis. There are more temporal cues representing joint events for this validation set.}
    \label{fig:val_temp_per}
\end{figure}

\begin{figure}
    \centering
    \includegraphics[width=0.98\columnwidth]{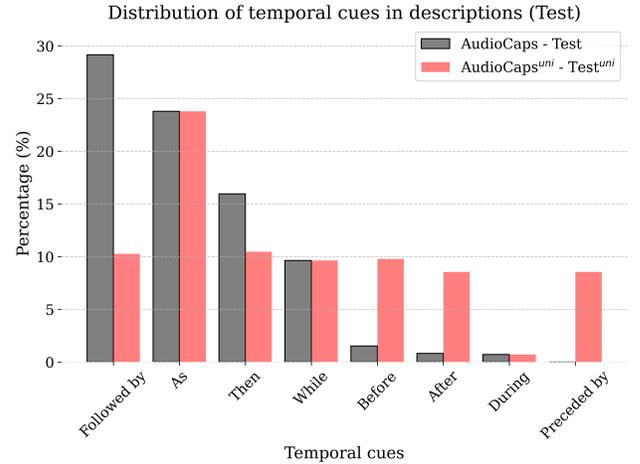}
    \caption{Distribution of temporal conjunctions and prepositions in the AudioCaps~\cite{kim2019audiocaps} test dataset (Test) compared to our proposed variant with a more uniform distribution of temporal textual cues ($Test^{uni}$). 
    }
      \Description{Grey and red histogram for the AudioCaps test set with percentages shown on the vertical axis and words being shown on the horizontal axis. There are more temporal cues representing future events for this test set.}
    \label{fig:test_temp_per}
\end{figure}

For our analysis, we construct a more \textit{uniform} version of the AudioCaps dataset with descriptions having a more balanced distribution of temporal conjunctions and prepositions. In particular, we rephrase AudioCaps descriptions to preserve the original meaning while varying the use of temporal conjunctions and prepositions. We investigate if this improves temporal understanding.
Specifically, we generate the \textit{$AudioCaps^{uni}$} dataset with corresponding \textit{$Train^{uni}$}, \textit{$Val^{uni}$} and \textit{$Test^{uni}$} subsets based on the AudioCaps dataset. We use two approaches to re-writing the descriptions. One is to replace the temporal cues with something that has the same meaning, e.g.\ `Bird singing followed by dog barking' is equivalent to `Bird singing before dog barking'. The second approach is to re-order the text location of events and also change the temporal cue, e.g.\ `Bird singing followed by dog barking' becomes `Dog barking after bird singing'. We present the distribution of temporal cues in the original AudioCaps dataset and its uniform variant in Fig.~\ref{fig:train_temp_per}, Fig.~\ref{fig:val_temp_per}, and Fig.~\ref{fig:test_temp_per}.

In addition to the more uniform AudioCaps train, validation and test sets, we create a test subset where at least one of the five text descriptions contains \textit{future} and \textit{past} temporal cues (as explained in Sec.~\ref{sec:audiocaps_dataset}). We do this to evaluate the performance of the model on sentences that actually contain temporal cues of interest. We call this subset \textit{TempTest}.\par
For Clotho, we use its original data in our experiments. Differently from our AudioCaps experiments, we do not create a dataset variant with more uniformly distributed temporal conjunctions, since the `future' and `past' cues only account for less than 15\% of the Clotho data. However, as for AudioCaps, we evaluate the model on multiple test subsets that allow us to analyse the temporal understanding capabilities.

\subsubsection{Experiments}
We consider three main experiments. 
First, we conduct the standard evaluation for text-to-audio retrieval on the AudioCaps and Clotho test sets. The performance on the standard test set serves as a point of reference for the training and evaluation on different variants of the data.

The second experiment involves reversing the ordering of sounds in the text queries of the test set. We refer to this as \textit{$Test^{rev}$}. The purpose of this experiment is to see what happens if the temporal text descriptions keep the same temporal preposition or conjunction but the sound sources are reversed, resulting in a wrongly ordered description, e.g.\ \textit{\underline{Birds singing} before \underline{dog barks}} becomes \textit{\underline{Dog barks} before \underline{birds singing}}. If the model understands the temporal ordering of events, the  model's performance should drop for the `wrongly' ordered events as compared to the original test set performance.

For the third experiment, 
we replace the temporal cue in a description with its opposite, thus changing the order of events without changing their spatial position in the text description, e.g.\ \textit{Birds singing \underline{before} dog barks} becomes \textit{Birds singing \underline{after} dog barks}. We refer to this as \textit{Test$^{rep}$}. More concretely, we make the following replacements: (`after’$\rightarrow$`before’), (`before’$\rightarrow$`after’), (`then’$\rightarrow$`before’), (`followed by’$\rightarrow$`preceded by’), and (`preceded by’$\rightarrow$`followed by’). \par 
If the model does not understand temporal cues, we expect it to perform similarly well on \textit{Test}, \textit{Test$^{rev}$} and \textit{$Test^{rep}$}. Conversely, if it understands temporal ordering, the \textit{Test} performance should be considerably higher than \textit{Test$^{rev}$} and \textit{Test$^{rep}$}. We would also expect  \textit{Test$^{rev}$} and \textit{Test$^{rep}$} to be similar, as the meaning of the sentence is the same but opposite of the correct \textit{Test} sentences. We additionally consider \textit{rev} and \textit{rep} subsets of the temporal subset 
 \textit{TempTest}. \par

In Tab.~\ref{tab:retrieval_results_origtrain_5_eval_3seed}, we take the checkpoint provided by~\cite{mei2023wavcaps} which was pre-trained on the WavCaps~\cite{mei2023wavcaps}, AudioCaps~\cite{kim2019audiocaps}, and Clotho~\cite{drossos2020clotho}, and finetune it on the original AudioCaps dataset (similar to~\cite{mei2023wavcaps}). 

We notice that on the reversed \textit{TempTest$^{rev}$} set the model performs worse, indicating that the model understands temporal ordering. However, on \textit{TempTest$^{rep}$} which contains the replacement of temporal cues, the model performs similarly to on \textit{TempTest}. This is interesting, as the temporal ordering of both \textit{$TempTest^{rev}$} and \textit{TempTest$^{rep}$} is reversed and wrong as compared to \textit{TempTest}. The only difference is that for the former, the actual positional text locations of the sounds are swapped, whilst for the latter the meaning is reversed by changing the temporal connector. This leads us to believe that at best, the model learns text location-based ordering rather than the ordering given by the text connector. In the left-hand side of Tab.\ref{tab:clotho_retrieval_results_combined} we observe a similar behaviour for the Clotho temporal test sets, with the model performing similarly well on \textit{TempTest$^{rep}$} and a bit worse on \textit{TempTest$^{rev}$}. This indicates that the model at best learns the text location-based ordering, but does not really understand temporal cues.

We do the same experiments on the test sets of AudioCaps$^{uni}$ and find that the model is unable to identify the text-based order of sound events, with results on `correct' (TempTest) and `wrong' (TempTest$^{rev}$ and TempTest$^{rep}$) splits being comparable. This is due to the test set not being biased anymore towards \textit{future} events. \par

Next, we examine whether the model's understanding of temporal ordering is limited due to a lack of variety in the training examples.
 We take the same pre-trained model as before~\cite{mei2023wavcaps}, and finetune it on the $AudioCaps^{uni}$ $Train^{uni}$ set. We notice that the overall performance on the \textit{$AudioCaps^{uni}$} test sets and the corresponding temporal subsets is higher when finetuning on a more uniform distribution of temporal cues (left half of Tab.~\ref{tab:retrieval_results_unitrain_audiocaps_combined}) than when finetuning on the original training data (bottom of Tab.~\ref{tab:retrieval_results_origtrain_5_eval_3seed}). Thus, the lack of understanding temporal ordering is in part due to the training data not containing examples of \textit{past} temporal cues. We also notice some signs of better temporal understanding, with a slightly bigger drop in performance on the \textit{TempTest$^{rev}$} and \textit{TempTest$^{rep}$} sets relative to \textit{TempTest}.\par

\begin{table}
\caption{Text-to-audio retrieval and audio-to-text retrieval on the AudioCaps and  AudioCaps$^{uni}$ datasets for the model fine-tuned on AudioCaps (Train). We report retrieval accuracies R@1. Reversing the order of events (\textit{rev, rep}) does not generally  result in a drop in performance, showing that the model does not understand temporal ordering. Lower performance on TempTest$^{rev}$ might be due to the bias induced by data being described in a \textit{future} fashion. When this bias is removed from the test set in AudioCaps$^{uni}$, reversing the order of events does not change performance.}
\centering
\resizebox{0.3\textwidth}{!}{
\begin{tabular}{cccc}
 \toprule
 Eval Dataset & \makecell{Subset}  & \multicolumn{1}{c}{T$\rightarrow$A} & \multicolumn{1}{c}{A$\rightarrow$T}\\
  \cmidrule(lr){3-4}
 & & R@1 & R@1 \\
 \midrule

\multirow{4}{*}{AudioCaps} & \makecell{Test} & 43.71 & 55.44 \\
& TempTest& 50.06  & 62.95   \\
&TempTest$^{rev}$ & 44.11 & 57.27\\
&TempTest$^{rep}$ & 49.12 & 63.12  \\
\addlinespace
\multirow{4}{*}{AudioCaps$^{uni}$} & \makecell{Test}&  41.61 & 53.80  \\
&TempTest& 48.04 & 61.74   \\
&TempTest$^{rev}$ & 47.29 & 62.29  \\
& TempTest$^{rep}$ & 47.24 & 62.23 \\

\bottomrule

\end{tabular}%
}
\label{tab:retrieval_results_origtrain_5_eval_3seed}
\end{table}

\begin{table}
\caption{Text-to-audio retrieval and audio-to-text-retrieval on the Clotho dataset for the model \textbf{fine-tuned on Clotho (Train)}. We report retrieval accuracies R@1. Reversing the order of events does not change performance when using only the text-audio contrastive loss. When training with the added text-text contrastive loss, evaluation on the wrong order of events (i.e.\ \textit{rev, rep}) yields a 
drop in performance.}
\centering

\begin{tabular}{ccccccc}
 \toprule
  Subset & Loss & \multicolumn{1}{c}{T$\rightarrow$A} & \multicolumn{1}{c}{A$\rightarrow$T} & Loss & \multicolumn{1}{c}{T$\rightarrow$A} & A$\rightarrow$T\\
  \cmidrule(lr){3-4}\cmidrule(lr){6-7}
& & R@1 & R@1 & & R@1 & R@1\\
 \midrule

\makecell{Test} &$\mathcal{L}_{ta}$ & 18.79 & 24.35 & $ +\lambda\mathcal{L}_{tt}$ & 18.17 & 21.56  \\
TempTest& $\mathcal{L}_{ta}$ & 30.84 & 39.33 &$+\lambda\mathcal{L}_{tt}$ & 30.02 & 36.24  \\
TempTest$^{rev}$ & $\mathcal{L}_{ta}$ & 30.02 & 38.48 & $+\lambda\mathcal{L}_{tt}$ & 24.57 & 34.93\\
TempTest$^{rep}$ & $\mathcal{L}_{ta}$ & 30.30 & 39.33  &  $+\lambda\mathcal{L}_{tt}$ & 24.34 & 33.52 \\

\bottomrule

\end{tabular}%

\label{tab:clotho_retrieval_results_combined}
\end{table}
\begin{table}
\caption{Text-to-audio retrieval and audio-to-text retrieval results on the temporally uniform AudioCaps$^{uni}$ dataset for the model fine-tuned on AudioCaps$^{uni}$ (Train$^{uni}$). Improved results on Test$^{uni}$. Using the text-text contrastive loss, the model better understands temporal ordering (larger drop for ${rev}$ and ${rep}$ compared to TempTest).
}
\centering

\begin{tabular}{ccccccc}
 \toprule
 Subset & \makecell{Loss}  & \multicolumn{1}{c}{T$\rightarrow$A} & \multicolumn{1}{c}{A$\rightarrow$T} & \makecell{Loss}  & \multicolumn{1}{c}{T$\rightarrow$A} & \multicolumn{1}{c}{A$\rightarrow$T}\\
  \cmidrule(lr){3-4}\cmidrule(lr){6-7}
& & R@1 & R@1 & & R@1 & R@1\\
 \midrule

 \makecell{Test}& $\mathcal{L}_{ta}$  & 43.52 & 53.40 &$+ \lambda\mathcal{L}_{tt}$ & 42.00 & 52.25 \\
TempTest& $\mathcal{L}_{ta}$& 50.71 & 62.41 &$+ \lambda\mathcal{L}_{tt}$ & 48.32 & 59.37 \\
TempTest$^{rev}$& $\mathcal{L}_{ta}$ & 46.82 & 59.43 &$+ \lambda\mathcal{L}_{tt}$ & 39.38 & 52.43\\
TempTest$^{rep}$& $\mathcal{L}_{ta}$ & 46.99 & 58.58  &$+ \lambda\mathcal{L}_{tt}$ & 38.42 & 51.52 \\

\bottomrule

\end{tabular}%

\label{tab:retrieval_results_unitrain_audiocaps_combined}
\end{table}

\section{Analysis of temporal understanding in controlled setting}\label{subsec:syncaps}
We analyse the text-audio model's temporal understanding in a controlled setting where we can guarantee correct alignment of text-audio pairs.

\subsection{Data generation} We use the ESC-50~\cite{piczak2015dataset} environmental sound classification dataset to generate a synthetic dataset for text-to-audio retrieval with a focus on temporal understanding capabilities. 
ESC-50 is a dataset of 2000 audio samples from 50 classes. As this dataset is clean and contains `atomic' sounds (i.e.\ 5 second audios containing only one sound), we use it for synthetic data generation.\par
We first task an LLM to take the sound labels from ESC-50 and generate textual descriptions in the style of AudioCaps (e.g.\ `dog'$\rightarrow`$dog barking'). To generate the text-audio pairs, we take two sounds and their LLM-generated labels and concatenate them based on an arbitrarily selected temporal order (see Fig.~\ref{fig:summary_figure} in the middle). We call this dataset \textit{SynCaps}.

To avoid any confusion, we only use \textit{future} and \textit{past} temporal cues. This is because synchronous temporal cues such as `as' or `during' are ambiguous, especially in a noisily labelled dataset. They can be used for sounds that completely overlap, or for partial overlaps of sounds, ignoring the actual order in which the sounds appear.
The test set contains unique sound components that are not used in the training and validation sets. This leads to 485 test examples of 10 second long audio clips.
For training and validation, we allow the same 5 seconds sound component to appear on average five times. We apply five different types of augmentation that include time shifting, volume adjustment, pitch shift, time stretch, and the addition of noise. We also allow for an overlap between the files of up to one second. This results in a total of 4400 training samples, 485 validation samples, and 485 test samples.

\subsection{SynCaps experiments}

We analyse the temporal understanding of the text-audio model in the controlled setting of the SynCaps dataset. For this, we take the same pre-trained model from~\cite{mei2023wavcaps} and finetune it on SynCaps using the original $\mathcal{L}_{ta}$ loss.

We observe that evaluating on the `reversed'(\textit{rev}) and `replaced' (\textit{rep}) datasets gives almost the same results as using the correct (original) test data (left half of Tab.~\ref{tab:retrieval_results_finetuned_temporal_newdataset_combined}). This shows that the model indeed does not understand temporal cues even on a simple dataset. \par

\section{Text-text contrastive loss}
We propose a loss function $\mathcal{L}_{tt}$ that aims to enhance the understanding of temporal information.
It is formulated as a text-text contrastive loss, which relies on pairs of positive examples (that have the same temporal significance as the original sentence) and negative text examples (that have the opposite temporal meaning). Concretely, given the original description \textit{Bird sings \underline{followed by} dog barks}, one positive example is \textit{Bird sings \underline{before} dog barks} and one negative example would be \textit{Bird sings \underline{after} dog barks}.

We provide the model with two positive text examples and two negative text examples for each text description containing the previously defined \textit{future} and \textit{past} temporal textual cues. Positive and negative text examples can be generated once, before training the model.
We searched for the temporal cues we are interested in and automatically generated multiple positives and negatives by changing the temporal cues and/or the ordering of the sounds.\par
The contrastive loss for each query and a margin $\alpha$ is:
\begin{equation}
\mathcal{L}_{tt} = \frac{1}{2N} \sum_{n=1}^{N} \sum_{k=1}^{2} \max(0, \alpha - s_{t_{n}{t_{k,pos}}} + s_{t_nt_{k,neg}}),
\end{equation}
where $s_{t_{n}{t_{k,pos}}}$ is the cosine similarity (see Eq.~\ref{eqn:sim}) between the $n$-th text query and its $k$-th positive example, $s_{t_{n}{t_{k,neg}}}$ is the cosine similarity between the $n$-th query and its $k$-th negative example.
Our full loss then becomes:
\begin{equation}
\mathcal{L} = \mathcal{L}_{ta} + \lambda\mathcal{L}_{tt}.
\end{equation}
In our experiments that use the text-text contrastive loss, we set $\lambda=10$, and $\alpha=0.2$.\par

\subsection{Performance using text-text contrastive loss}
We evaluate the same model pre-trained on the joint WavCaps $+$ AudioCaps $+$ Clotho, and finetuned on SynCaps, AudioCaps$^{uni}$ and Clotho using our additional text-text contrastive loss. For SynCaps, we observe in Tab.~\ref{tab:retrieval_results_finetuned_temporal_newdataset_combined}, that the model performs better on the original test set when the additional text-text contrastive loss is used, whilst at the same time showing a big drop in performance on the `reversed' and `replaced' data. This shows that employing a simple additional loss can help the model better understand time, at least in the controlled setting of the SynCaps dataset.\par
\begin{table}
\caption{Text-to-audio and audio-to-text retrieval on the SynCaps dataset for the model fine-tuned on SynCaps with the text-audio loss $\mathcal{L}_{ta}$, and with the text-text loss $\mathcal{L}_{tt}$. When using only $\mathcal{L}_{ta}$, there is almost no drop in performance on the wrongly ordered test sets $rev$, $rep$. With the additional $\mathcal{L}_{tt}$, the drop becomes much more significant, confirming that the model understands temporal ordering.}
\centering
\begin{tabular}{ccccccc}
 \toprule
 \makecell{Subset} & Loss  & \multicolumn{1}{c}{T$\rightarrow$A} & \multicolumn{1}{c}{A$\rightarrow$T} & Loss & \multicolumn{1}{c}{T$\rightarrow$A} & \multicolumn{1}{c}{A$\rightarrow$T}\\
  \cmidrule(lr){3-4}\cmidrule(lr){6-7}
& & R@1 &  R@1 & & R@1 &  R@1\\
 \midrule

 \makecell{Test} & $\mathcal{L}_{ta}$ & 67.22 &65.23 & $+ \lambda\mathcal{L}_{tt}$  & \textbf{69.35}  & \textbf{69.90} \\
\makecell{Test$^{rev}$} & $\mathcal{L}_{ta}$ & 67.35 & 65.84 & $ + \lambda\mathcal{L}_{tt}$ & 40.55  & 41.03\\
   \makecell{Test$^{rep}$} & $\mathcal{L}_{ta}$ & 66.94 & 63.92 & $+ \lambda\mathcal{L}_{tt}$ & 44.33  & 45.43  \\

\bottomrule

\end{tabular}%

\label{tab:retrieval_results_finetuned_temporal_newdataset_combined}
\end{table}

When finetuning the WavCaps-pretrained model on AudioCaps$^{uni}$ with the text-text contrastive loss, we see in the right-hand side of Tab.~\ref{tab:retrieval_results_unitrain_audiocaps_combined} that the drop is larger between the TempTest and the ${rev}$ and ${rep}$ test sets than when just using the text-audio original loss function. At the same time, the performance on the full Test$^{uni}$ and TempTest remains competitive when using the additional loss as compared to only using the original one.\par

Lastly, we show that on the original form of the Clotho dataset, when finetuning using the text-text contrastive loss, we obtain a bigger drop on the ${rev}$ and ${rep}$ test subsets. This indicates that although Clotho contains considerably more simultaneous actions and the number of \textit{future} and \textit{past} temporal cues is greatly disproportionate, the loss still helps the model better focus on temporal cues when present. This is observed in Tab.~\ref{tab:clotho_retrieval_results_combined}.

\section{Conclusion}
In this work, we dissected the temporal understanding capabilities of a current state-of-the-art text-audio model. We first performed an in-depth analysis of the AudioCaps and Clotho text-audio datasets. We concluded that these datasets are not well-suited for obtaining and evaluating temporal understanding capabilities in a text-audio retrieval model. As a result, we proposed a variant of the AudioCaps dataset, namely AudioCaps$^{uni}$, that contains a more uniform distribution of different temporal cues. We then showed that using AudioCaps$^{uni}$ reduces biases learnt by the model and improves performance on the AudioCaps$^{uni}$ test set. Furthermore, we introduced a synthetic dataset (SynCaps), showing that indeed models fail to use the temporal cues even in a controlled data setting. Lastly, we proposed a simple loss that results in better text-to-audio retrieval results on SynCaps, whilst also putting more emphasis on the temporal content of the audio and text data in all datasets analysed.

\begin{acks}
This work was supported by an EPSRC DTA Studentship, by the Royal Academy of Engineering (RF\textbackslash201819\textbackslash18\textbackslash163), by the DFG: SFB 1233, project number: 276693517, and by the DFG EXC number 2064/1 – project number 390727645. We are very grateful to Samuel Albanie and Bruno Korbar for helpful feedback and suggestions. 
\end{acks}
\clearpage

\bibliographystyle{ACM-Reference-Format}
\balance
\bibliography{new_bib}


\end{document}